\documentclass[runningheads,a4paper]{llncs}

\usepackage[T1]{tipa}
\usepackage[english]{babel}
\usepackage{courier}
\usepackage{amstext}
\usepackage{times}
\usepackage{helvet}
\usepackage{courier}
\usepackage[cmex10]{amsmath}
\usepackage{amssymb}
\usepackage{latexsym}
\usepackage{eurosym}
\usepackage{wasysym}
\usepackage{array}

\makeatletter
\newif\if@restonecol
\makeatother

\usepackage{etex}
\usepackage{tikz}
\usepackage[T1]{fontenc}
\usepackage{pslatex}
\usepackage[arrow, matrix, curve]{xy}
\usepackage{epsfig}
\usepackage{pstricks,egameps}
\usepackage{hyperref, url}
\usepackage[algochapter, boxruled, linesnumbered]{algorithm2e}

\begin{document}

\title{What Government by Algorithm Might Look Like}
\titlerunning{What Government by Algorithm Might Look Like}

\author{Rustam Tagiew\inst{1}}
\authorrunning{Rustam Tagiew}

\institute{Alumni of TU Bergakademie Freiberg and University of Bielefeld\\
\email{rustam@tagiew.de}, \texttt{www.tagiew.de}}
\maketitle

\begin{abstract} 
Algocracy is the rule by algorithms. This paper summarises technologies useful to create algocratic social machines and presents idealistic examples of their application. In particular, it describes smart contracts and their implementations, challenges of behaviour mining and prediction, as well as game-theoretic and AI approaches to mechanism design. The presented idealistic examples of new algocratic solutions are picked from the reality of a modern state. The examples are science funding, trade by organisations, regulation of rental agreements, ranking of significance and sortition. Artificial General Intelligence is not in the scope of this feasibility study.  
\end{abstract}
\section{Introduction}
\indent During the COVID-19 pandemic in spring of 2020, the phenomenon of the rise in popularity of the 'human microchip implant' conspiracy theory could be observed, which has its roots in the biblical 'mark of the beast' prophecy \cite{markofbeast}. Despite the missing evidence, its worldwide popularity growth reveals the existence of public apprehension about the possible radical transformation of forms of government and social orders due to the progress in information technology. This type of fear previously known from cyberpunk literature even motivated numerous worldwide arson attacks on 5G towers.\\
\indent According to a 2019's poll of IE University of Spain, only $25$\% of European citizens are in favour of letting algorithms run their countries \cite{iestudy}. The Netherlands showed the highest approval rate of $45$\% in this study. Nevertheless, the algorithmic predictor of welfare fraud SyRI caused a public protest in the same Netherlands in the same year \cite{syri}. On the other hand, since 2012 secretive deployments of algorithms in public regulation like the cooperation between Palantir Technologies and New Orleans Police Department are documented \cite{palantir}. The use of information technology in public regulation is connotated with conspirative actions of the minority without consent of the majority. This might encumber further progress of such technologies for the common good. The apprehension of a radical transformation should be addressed in open scientific research to clear the fog of superstition and conspiracy theories, to improve technologies and to identify possible advantages.\\
\indent Application of information technology in governmental administration enabled E-Government, which is more efficient communication within state agencies, between state and non-state organisations, and between state and citizens. But, the actual radical transformation will be the decision-making by algorithms in place of humans and is not covered by the term E-Government \cite{egovalgo}.\\
\indent The rule by algorithms can be termed as algocracy, where algo- is derived from algorithm and -cracy is 'rule' in ancient Greek. Unfortunately, academic literature creates no consensus in the usage of any single term. A. Aneesh introduced the term {\it algocracy} in 2006 \cite{aneesh}, T. O'Reily introduced the term {\it algorithmic regulation} in 2013 \cite{toreilys} and Stanford's 2020's report about the application of AI in US government agencies uses the term {\it government by algorithm} \cite{govbyalgorithm}. Other terms for the same issue are {\it algorithmic power} \cite{beeralgo}, {\it governing algorithms} \cite{governingalgos} and {\it algorithmic governance} \cite{algogovernance}. The term {\it cybercracy} decodes as rule by effective use of information, which does not necessarily imply the usage of algorithms \cite{cyberocracy}. This paper will use the term algocracy.\\
\indent Algocracy is not a basic form of government, it is an enhancement to the existing forms of governments. Algorithms are not living beings and have no intrinsic preferences and goals unless introduced by human engineers. T. Berners-Lee describes processes, which combine people and algorithms as social machines \cite{Shadbolt2016}. In social machines, people are coordinated and administrated by algorithms and algorithms are engineered by people \cite{socialmachine}. Surely, only a small group of people will be competent enough to engineer and maintain the algorithms. In an algocratic dictatorship, the {\it preferences for algocratic social machine} (PASM) would come from the dictator and in an idealistic algocratic democracy from the people. However, a study of the real democracy in the United States showed that adopted policies are most correlated with the economic elites’ preferences at $0.79$ and have no significant correlation to the preferences of common citizens \cite{gilens_page_2014}. There is also some weak correlation to the preferences of business and mass-based groups. It is a question, whether an algocratic democracy will replace economic elites by engineering elites as the major source of preferences for the adopted policies, or the economic elites will command the engineers, or both elites' types will merge.\\
\indent Dictatorial states might have only one faction of these engineers and pluralist states multiple factions. These factions might compete peacefully against each other as companies on the free market, as political agendas in democratic elections or as open-border states in foot voting. Non-peaceful competition would include espionage, cyber warfare, economical warfare and violent hostilities. According to Y. N. Harrari, democracies outperformed dictatorships concerning innovation and economic growth in the late 20th-century, because they were better at processing information, but will underperform against dictatorships using central algorithms in the future \cite{harrari}. It can also be argued that an algocratic dictatorship might lose its technological edge over time, if it only has one faction of algocracy engineers without internal competition. Algocracy can be studied without predictions about the source of PASM. Let us only assume that PASM are always there no matter they came from a single dictator, certain elites or people.\\
\indent There is one important difference to traditional governments -- algocracy is constrained to have a certain minimal level of technology and industrialisation for production of data centres, communication infrastructure, surveillance equipment, etc.. A sustainable algocracy, therefore, requires a sustainable technosphere of certain minimal ecological footprint. This ecological footprint assembles as the space needed for the production of energy from sustainable sources as well as the space needed for the manufacturing and recycling facilities. Given a planet of finite space, algocracy is also constrained by a certain maximal expansion of the technosphere, which can be safely taken without seriously damaging the biosphere. These two boundaries make the PASM more predictable wherever they came from.\\
\indent Algocracy has to coordinate technosphere and the human population. Coordination of technosphere is a trivial command and control system. Coordinating human population is highly non-trivial and studied by multiple disciplines including jurisprudence and behaviour science. Every single human is an individual with her or his preferences, faults and thinking patterns. There are many ways to achieve coordination and the most obvious way is the formation of organisations. According to organisational theorist A. Stinchcombe, organisations can be viewed as contracts or complexes of contracts \cite{Stinchcombe}. The contracts like rental agreement, trade deals, employment contracts and so on are enforced by the judiciary system of a state. And the state including its judiciary system itself can be viewed as a social contract according to J.-J. Rousseau \cite{rousseau}. Seeing citizenship and state laws as contractual agreements even experience modern revival in the form of free private cities initiatives \cite{rahim}. In the context of free private cities, a state is regarded as a {\it government service provider}, which concludes a contract with its citizens \cite{titusgebel}.\\
\indent This paper's goal is to investigate setting up and enforcement of complexes of contracts as the part in the coordination of population by the means of algorithms. This impacts the three forces of powers, the legislature, the judiciary and the executive. Legislature and judiciary are considered to be fully automated using Weak AI in the near future. The full automation of the executive requires Artificial General Intelligence (AGI). For instance, a fully-automated ability to react to unexpected socioeconomic crises like the COVID-19 pandemic is beyond the definition of Weak AI. Government by AGI is highly controversial and not relevant in this paper.\\ 
\indent The next section \ref{milestones} describes the historical development. Section \ref{weakaialgocracy} makes a summary of relevant technologies. Then in section \ref{examples}, five examples of algocratic solutions to common problems are presented and a conclusion in section \ref{conclusion} lists seven theses as a result.\\ 
\section{Historical Milestones}
\label{milestones}
\indent Contracts, which are enforced by machines instead of state agencies, judges and investigators, are called smart contracts \cite{smartcontract}. A vending machine is the first simplest piece of technology known to humanity since the times of Heron of Alexandria, which manifests a smart contract \cite{Stinchcombe}. Proposals for coordination of economy by algorithms are known since the second half of the 20th century. A. Kharkevich proposed processing information in order to control their economy in 1962 \cite{khar} and Chile deployed {\it Project Cybersyn} in 1971--1973 \cite{cybersyn}.\\
\indent Since the 1960s, H. Simon pioneered expert systems for judiciary and administration \cite{herbertsimon}. Such branches of administration as tax offices have a decades-long history of the deployment of rule-based systems \cite{expertsystems}. Examples of those automated systems for legal reasoning are TAXMAN by T. McCarty and LEGOL by R. Stamper \cite{taxman,legol}.\\
\indent The term smart contract was introduced by N. Szabo in 1994 \cite{szabo}. Since 2000s, systems for automated surveillance are developed \cite{autsurveilance}. China started to develop its Social Credit System from 2009 on \cite{socialcreditsystem}. Social Credit System is actually a smart contract about the calculation of certain reputation score based on information available to the system from mass surveillance. The calculated score has far-reaching consequences on the lives of Chinese citizens -- it automatically improves or denies access to a broad category of services. Multiple non-state facilitators of platform economies like Uber or AirBnB employ less restrictive reputation systems, where interaction peers take decisions based on public reputation records rather than a central authority doing it.\\
\indent Since 2014 smart contracts were redefined more precisely as contracts, whose enforcement also does not require any trusted computing \cite{buterin}. In the newer sense, smart contracts are programs, which are executed on a blockchain virtual machine and can not be manipulated by any node in a solo attempt. The terms of a smart contract are the code of the smart contract. In 2016, the first decentralised autonomous organisation {\it The DAO} was launched as a smart contract on a blockchain, which resembled an investor-driven venture capital fund \cite{thedao}.\\
\indent The automation of judiciary system also progressed in recent years. Since 2016, multiple US courts use COMPAS software, which assists in risk assessment of recidivism prior to a court decision \cite{compas}. China claims to be the first country to run an AI automated court in 2019 \cite{cybercourt}. Also, Estonia uttered similar plans of letting AI decide on cases under \EUR{$7.000$} \cite{estonia}. Automation of judiciary leads to the same result as the replacement of judiciary by self-executing smart contracts.\\
\section{Weak AI Algocracy}
\label{weakaialgocracy}
\begin{figure}[t]
\begin{center}
\includegraphics[scale=0.55]{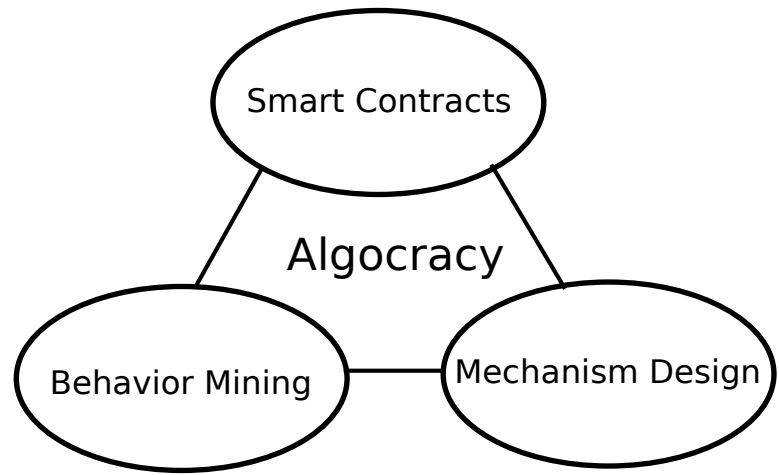}
\caption{The triangle of research fields for Weak AI algocracy.}
\label{algocracy}
\end{center} 
\end{figure}
\indent Fig.\ref{algocracy} depicts the main technology areas relevant to algocracy. Smart contracts are not only contracts between individuals and companies. Smart contracts are also laws, which apply to state and citizens. We speak about self-executing laws, which include crime regulations, taxation and obligations of the state. For our vision of algocracy, the execution of smart contracts has to be trustful, but their execution on a blockchain is not obligatory. In fact, a blockchain implementation can not hinder the communication of contract participants over additional channels, if it is required by the contract. A blockchain solely provides trustful computation, even if none of the nodes can be trusted.\\
\indent The process of engineering of new smart contracts falls into the discipline of mechanism design. As already been assumed, the PASM are always present and the mechanism design needs some assumptions about preferences and behaviour patterns of the human population. Obviously, there is a large amount of smart contracts, which will undermine the PASM, even if the participation in them is voluntary. Such harmful contracts have to be blocked. The harm from a smart contract can range from functional bugs like in the famous Ethereum hack of 2014 to more subtle consequences like in case of a bogus reputation system e.g.. The engineering of smart contracts has the goal to achieve the predefined PASM by useful contracts and to block harmful contracts.\\
\indent Mechanism design is originally a sub-field of game theory, but also can be seen as a sub-field of AI. Mechanism design should produce rules to achieve certain behaviour of the players, once preferences of interacting individuals are known. It is a reverse task to the rest of game theory, where the rules are known and the behaviour of the players aka Nash equilibrium is wanted. Classical game theory research came up with homo economicus assumption and saw humans being exclusively interested in monetary incentives, rational and able to foresee the consequences of their action many turns ahead. Field and laboratory data extenuated this assumption and revealed a more complicated structure of incentives, boundaries to rationality and ability to reason. Assumption of rationality is only a useful orientation in the analysis of behavioural data, which helps to reduce the hypotheses space \cite{osipov2019neural,kadyrov2019attribution,giftexchange}.\\ 
\begin{figure}
\begin{center}
\includegraphics[scale=0.4]{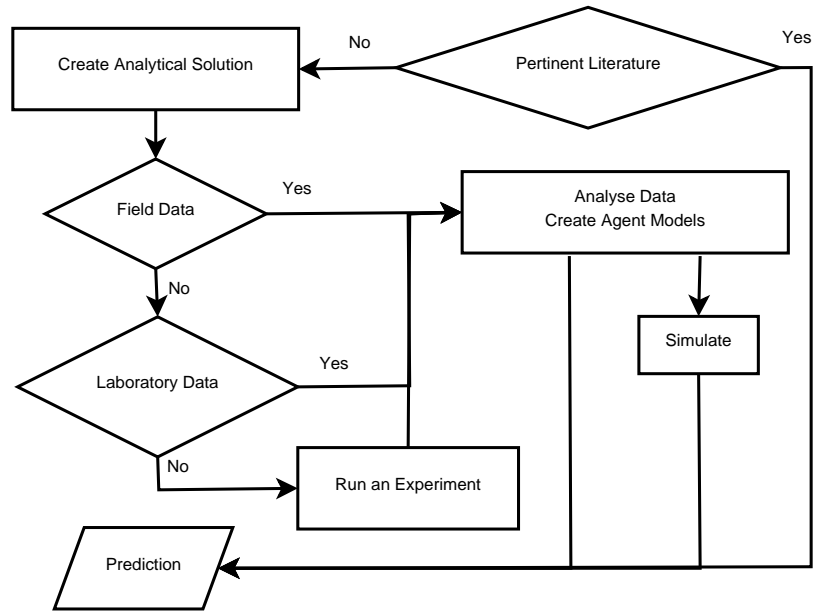}
\caption{Algorithm for prediction of consequences of a given smart contract.}
\label{prediction}
\end{center} 
\end{figure}
\indent In a non-algocratic society, a process related to mechanism design takes either place in a democratic parliament or behind closed doors in a less democratic government form. For instance, laws are introduced, which discourage crime by punishment. This process is crucial for the survival and success of the whole society. Given that the pursuit of certain PASM are taken seriously, the creation of appropriate smart contracts will be a non-trivial process and requires major complex calculation indeed beyond human capabilities, but still in the tangibility for Weak AI.\\
\indent Mechanism design can be seen as a search in the space of possible smart contracts, which is performed by genetic programming e.g.. In order to evaluate the gain or the harm from a smart contract, a prediction of its consequences has to be calculated. Fig.\ref{prediction} shows an abstract definition of the algorithm for this prediction. If there is any pertinent literature or any other kind of a legitimate record about the consequences, the solution is found.\\
\indent If there is no literature, an analytical solution has to be created to guide the data analysis. This can be done by game-theoretic or AI algorithms. Field data of human behaviour produced by the actual or similar contracts would be ideal for the analysis, even if it is noisy. The field experimental economics provides a certain amount of clean data sets, which can help to predict human behaviour if field data is not available. Finally, if no useful data is available, a field or laboratory experiment should be run. The goal is either to derive a prediction or to build agent models of human behaviour using this data. And this task is called behaviour mining or also cognitive modelling. If the analysis of the data is not sufficient to make a prediction, a simulation with agent models of human behaviour has to be used. The whole algorithm of engineering a useful smart contract does not include any parts, which require AGI.\\ 
\indent The core element of this vision of algocracy is an appropriate scripting language for smart contracts. Scripting languages for blockchains are a special issue. There is a discussion about the need for Turing-completeness. In 2019, only $35,3$\% of $53757$ smart contracts from Ethereum included loops and recursions \cite{jansen2019smart}, which are related to the halting problem. The Turing-completeness of the language creates more risk for bugs. The language should be of maximal usability and readability for humans. Algorithms for analytical solutions, tools for experiments, data analysis engines and simulation environments should understand the same language without reformatting. Today, both types of scripting languages exist -- with and without Turing-completeness.\\
\indent Simulation of agent interactions falls in the scope of General Game Playing (GGP). GGP is a design of AI algorithms, which can not only interact according to some special rules but also understand any rules written in special language \cite{pellfirst}. It is a step towards AGI -- design of general algorithms, which can play chess, checkers and poker likewise. And also for GGP, there exist two types of language. PNSI is not Turing-complete, because it is based on Petri Nets \cite{tagiew2008multi}. GDL is Turing-complete and is based on Datalog \cite{saffidine2014game}. Scripting languages for experimental economics' tools like z-Tree are obviously Turing-complete \cite{zTree}.\\
\indent In game theory, the most powerful class of games are games of imperfect information with moves of nature. Moves of nature are random events. Imperfect information is missing knowledge about other players' turns or random events. Fig.\ref{renting} shows an example of such a game. The realisation of random number generator on a blockchain is a complicated issue since the computation has to deliver the same result independent of the node performing it. This issue can be solved, if a source of entropy can be introduced, which can not be manipulated by participants \cite{chatterjee2019probabilistic}. Such sources of entropy can be the timestamps, transaction statistics, numbers committed by participants, smart contracts and sequences of actions from mixed strategy Nash equilibria games. Imperfect information is a more complicated issue. For instance, a Vickrey auction delivers the auctioneer optimal results, only if the bidders do not see other bids and do not communicate. Even if a Vickrey is reliably implemented inside of a blockchain, the communication of the bidders can not be prevented, if they use other communication channels.\\
\section{Examples}
\label{examples}
\subsection{Renting Agreement}
\begin{figure}[h]
\begin{center}
\includegraphics[scale=0.45]{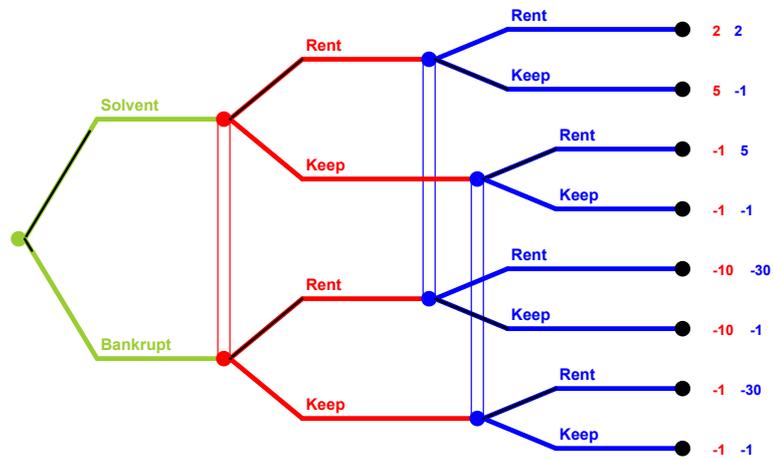}
\caption{Big land-lords' (red) and small land-lords' (blue) game. Green lines depict the probability of a tenant being solid or bankrupt in future. Narrow lines connect nodes indistinguishable for a particular decision-maker. Black lines depict the calculated Nash equilibrium.}
\label{renting}
\end{center} 
\end{figure}
\indent Modern economies rely on the workforce, which can relocate according to workforce demand. It is therefore important, that the employees can find affordable shelters. But, tenants are prone to bankruptcy and bankrupt tenants are prone to homelessness. Homelessness often causes irreversible damage to the affected people and therefore damages the whole society. If a tenant is bankrupt and can not pay the rent, the landlord is interested in a speedy eviction. A bankrupt tenant is interested in delaying eviction as much as possible.\\
\indent In Germany for instance, a delay of two months' rents legitimates the landlord to take legal actions to achieve an eviction. In reality, the landlord waits longer than a year until the eviction can be executed by authorities. The landlord bears the costs and has to suffer possible vandalism. Renting loss insurances only incur renting contracts with tenants of clearly verifiable income statements and credit scores. Due to higher financial reserves, big landlords do not need this insurance at all. In case of tenants with less clear scores and income statements, a small landlord is more affected than a bigger one, since he is more prone to financial risks and management mistakes.\\
\indent Fig.\ref{renting} shows the game-theoretic representation of this interaction. For small land lords, the described regulation can render the renting unprofitable for tenants of low credit score. This reduces the mobility of young employees of lower credit scores in return since they get less offers. On the other hand, the big landlords can demand higher rents. Algorithmic legislation and law enforcement might be very useful for automatic optimisation of the regulation depending on the economic situation. It could also evaluate and introduce some additional measures like an obligatory renting loss insurance.\\   
\subsection{Trade by Organisations}    
\indent In 2017, Ukraine's justice ministry conducted trial auctions using blockchain technology to improve transparency in governmental transactions \cite{ukraine}. Transparency is needed to hinder corruption through collusion in auctions of governmental assets -- the unlawful benefits of public officials. If PASM require deals to be sealed at best conditions possible for the whole social machine and not for some individuals, trustfully conducted auctions offer a mechanism to achieve it. Corruption in trade by organisations does not only cause damage on the state level, but also on the corporate level. Procurement in supply chains requires constant optimisation by competition of suppliers not only regarding costs, but also secondary criteria as ecological footprint, work conditions in production and so on. These secondary criteria are important for the prosperity of the whole social machine beyond the goal of the single enterprises sealing deals in a supply chain.\\
\indent Developing auction rule sets is not trivial and has a long history. Already 500 BC in Babylon, maidens were auctioned in forward and reverse auctions \cite{herodotus}. In a forward auction, the bride price to be paid by the swains is determined, while a reverse auction is about the dowry to be paid to the swains. Since then, the auctions, fortunately, moved into less barbaric contexts and the number of used rule sets for them increased dramatically \cite{friedrich}. With the rise of the internet, more complicated auction rule sets could be implemented, which were unfeasible without computer systems. For instance, a multi-attribute auction allows bids, which consist of more than one attribute \cite{multiattribute}. In a standard English auction, a bid is a price, which has to be higher than the previous bid in forward auction case and lower in a reverse auction case. In a multi-attribute auction, the price will be only one attribute of many and a utility function determines, whether a bid is an improvement upon the previous one. Special auction rule sets can be optimised and deployed in diverse contexts inside of an algocratic social machine. The goal is not to command free enterprises and free individuals, but to set the right incentives for the common goals.\\ 
\subsection{Research Founding}    
\indent Industrialised countries spend single-digit perceptual shares of GDP on R\&D. The optimisation of technosphere required for algocracy relies on efficient research. Research funding has to be spent on the most promising projects and most promising individuals. Currently, popular scientometric measures like $h$-index set disputable incentives \cite{tagiew2017}. For instance, the growth of the author's number per paper is incentivised, since increased the number of papers per author and also the number of citations. The data of scientific publications shows a clear proof that this incentive makes an impact. Further, it incentivises organisational and political talents instead of actual innovation originators. An algocratic social machine hands over a lot of organisational and political tasks to algorithms and therefore will concentrate the funding on innovation originators. Technologies like text mining and data citation could play a crucial role in this process.\\  
\subsection{Attention Inequality}
\indent There is a silent agreement between a search engine provider and its user -- search engine provides most relevant results on top to any search request given by the user. This agreement is not a formal contract and does not involve the actual third party namely the search result providers. A similar agreement between major media outlets and their readers. The supposedly most important stories are on the title page. Also for media, these agreements did not include any third parties till now. Finally, social networks also apply certain filtering on user-generated contents based on some significance ranking \cite{attenineq}.\\
\indent Given the fact that attention is a scarce asset on the internet and a precondition for monetary income and other non-monetary benefits, a justification for this attention inequality becomes an issue of interest. In the case of media, bizarre effects like {\it missing white woman syndrome} and {\it hierarchy of death} can be observed \cite{missingwhitewoman,hierarchyofdeath} -- people's suffering gets different coverage depending on their origin. This is clearly a dysfunction. In less critical contexts, attention inequality achieves generally much higher GINI coefficients than wealth inequality \cite{attengoog}. For an algocratic social machine, this will be important to assess the justified share of attention inequality. The unjustified share of attention inequality can be explained by Matthew effect, which can make a low-quality content to get more attention than a high-quality content \cite{matthew}. 
\subsection{Sortition}    
\indent Sortition is the selection of political officials by random, which is known since ancient Athenian democracy. According to the law of large numbers, sortition would ensures the implementation of the average will of the general public for larger pools of selected officials. For instance, if sortition creates a relatively small random parliament of $22$ seats from a population of $1$M people including $50$k proponents of a dictatorship, the probability of these radicals getting $11$ seats and more is only $0.000000002$ -- it will take millions of years until the democracy is overthrown by chance. Sortition appears to be more secure at fulfilling democracy than an election. As already mentioned, the political decisions in the US democracy do not correlate with the will of the general public \cite{gilens_page_2014}. Otherwise, political parties are reported to spend significant amounts of money on data-driven electoral campaigns \cite{hankey2018data} -- voters' behaviour is influenced by algorithms. Ancient Greeks built special machines to ensure a fair random selection. In algocracy this task will be done by algorithms.\\
\section{Conclusion}
\label{conclusion}
\indent We made a summarising journey from cloudy fears to the main technologies and concepts around the coming radical transformation by algocracy. Main theses are:
\begin{enumerate}
\item Historical milestones show clear worldwide trends towards establishment of algocracy regardless of political background. 
\item Majority of the population is not in favour of this transformation, although algocracy has the potential to address such problems as environmental degradation, discrimination and unjustified economic inequalities.
\item There are no clear signs that algocracy might shift the power balance either towards even smaller elite or towards the populace.
\item The legislature and the judiciary can be automated by Weak AI and full automation of the executive requires AGI.
\item Automation of judiciary requires smart contract implementation.
\item Automation of legislature is a data-driven process of mechanism design based on simulation.
\item Development of a language for smart contracts is the core element for algocracy since it builds the interface between the virtual machine, the mechanism design, the data formatting and the human subjects.\end{enumerate}
\bibliographystyle{splncs}
\bibliography{algocracy}

\begin{thebibliography}{10}

\bibitem{markofbeast}
{Wired Staff}:
\newblock {RFID: Sign of the (End) Times?}
\newblock Wired.com (2006)

\bibitem{iestudy}
Alcazar, D., Muniz, M.:
\newblock European tech insights.
\newblock Center for the Governance of Change, IE CGC (2019)

\bibitem{syri}
Simonite, T.:
\newblock {Europe Limits Government by Algorithm. The US, Not So Much}.
\newblock Wired (2020)

\bibitem{palantir}
Winston, A.:
\newblock {Palantir has secretly been using New Orleans to test its predictive
  policing technology}.
\newblock The Verge (2018)

\bibitem{egovalgo}
Veale, M., Brass, I.:
\newblock Administration by Algorithm? Public Management Meets Public Sector
  Machine Learning.
\newblock In: Algorithmic Regulation. Oxford University Press (2019)

\bibitem{aneesh}
Aneesh, A.:
\newblock Virtual Migration.
\newblock Duke University Press (2006)

\bibitem{toreilys}
O{'}Reilly, T.:
\newblock Open Data and Algorithmic Regulation.
\newblock In: Beyond Transparency: open Data and the Future of Civic
  Innovation. Code for America Press (2013)  289--300

\bibitem{govbyalgorithm}
Engstrom, D.F., Ho, D.E., Sharkey, C.M., Cuellar, M.F.:
\newblock Government by algorithm: Artificial intelligence in federal
  administrative agencies.
\newblock Stanford Law School (2020)

\bibitem{beeralgo}
Beer, D.:
\newblock {Power through the algorithm? Participatory web cultures and the
  technological unconscious}.
\newblock New Media and Society \textbf{11}(6) (2009)  985--1002

\bibitem{governingalgos}
Barocas, S., Hood, S., Ziewitz:
\newblock Governing algorithms: A provocation piece.
\newblock In: Governing Algorithms: A conference on computation, automation and
  control. (2013)

\bibitem{algogovernance}
Williamson, B.:
\newblock Decoding identity: Reprogramming pedagogic identities through
  algorithmic governance.
\newblock In: British Educational Research Association conference, University
  of Sussex (2013)

\bibitem{cyberocracy}
Ronfeldt, D.:
\newblock {Cyberocracy, Cyberspace, and Cyberology: Political Effects of the
  Information Revolution}.
\newblock RAND Corporation (1991)

\bibitem{Shadbolt2016}
Shadbolt, N., Kleek, M.V., Binns, R.:
\newblock The rise of social machines: The development of a human/digital
  ecosystem.
\newblock IEEE Consumer Electronics Magazine \textbf{5} (4 2016)  106--111

\bibitem{socialmachine}
Cristianini, N., Scantamburlo, T.:
\newblock On social machines for algorithmic regulation.
\newblock AI \& Society (2019)

\bibitem{gilens_page_2014}
Gilens, M., Page, B.I.:
\newblock Testing theories of american politics: Elites, interest groups, and
  average citizens.
\newblock Perspectives on Politics \textbf{12}(3) (2014)  564--581

\bibitem{harrari}
Harari, Y.N.:
\newblock Why technology favors tyranny.
\newblock The Atlantic (2020)

\bibitem{Stinchcombe}
Savelyev, A.:
\newblock Contract law 2.0: "smart" contracts as the beginning of the end of
  classic contract law.
\newblock Social Science Research Network (2016)

\bibitem{rousseau}
Rousseau, J.J.:
\newblock Du contrat social; ou, Principes du droit politique.
\newblock AMSTERDAM (1762)

\bibitem{rahim}
Taghizadegan, R.:
\newblock The Politics of Seasteading.
\newblock In: Seasteads. Opportunities and Challenges for Small New Societies.
  vdf (2017)  67--79

\bibitem{titusgebel}
Gebel, T.:
\newblock Freie Privatstädte.
\newblock Aquila Urbis (2018)

\bibitem{smartcontract}
Röscheisen, M., Baldonado, M., Chang, K., Gravano, L., Ketchpel, S., Paepcke,
  A.:
\newblock The stanford infobus and its service layers: Augmenting the internet
  with higher-level information management protocols.
\newblock In: Digital Libraries in Computer Science: The MeDoc Approach.
\newblock Springer (1998)  213--230

\bibitem{khar}
Kharkevich, A.:
\newblock {Informatsia i Tekhnika}.
\newblock Communist \textbf{17}(94) (1962)

\bibitem{cybersyn}
Medina, E.:
\newblock Rethinking algorithmic regulation.
\newblock Kybernetes \textbf{44}(6/7) (2015)  1005--1019

\bibitem{herbertsimon}
Simon, H.A.:
\newblock Administrative Behavior.
\newblock (1997)

\bibitem{expertsystems}
Margretts, H.:
\newblock Information technology in government.
\newblock Routledge (1999)

\bibitem{taxman}
McCarty, T.L.:
\newblock Reflections on taxman: An experiment in artificial intelligence and
  legal reasoning.
\newblock Harvard Law Review (1977)  837--893

\bibitem{legol}
Stamper, R.K.:
\newblock The legol 1 prototype system and language.
\newblock The Computer Journal \textbf{20} (1977)  102--108

\bibitem{szabo}
Szabo, N.:
\newblock Smart contracts.
\newblock szabo.best.vwh.net (1994)

\bibitem{autsurveilance}
Sodemann, A.A., Ross, M.P., Borghetti, B.J.:
\newblock A review of anomaly detection in automated surveillance.
\newblock IEEE Transactions on Systems, Man, and Cybernetics, Part C
  \textbf{42}(6) (2012)  1257--1272

\bibitem{socialcreditsystem}
Zhong, Y.:
\newblock Rethinking the social credit system: A long road to establishing
  trust in chinese society.
\newblock In: Symposium on Applications of Contextual Integrity. (2019)  28--29

\bibitem{buterin}
Buterin, V.:
\newblock A next-generation smart contract and decentralized application
  platform.
\newblock WhitePaper (2014)

\bibitem{thedao}
Waterss, R.:
\newblock Automated company raises equivalent of 120m in digital currency.
\newblock Financial Times (2016)

\bibitem{compas}
Kirkpatrick, K.:
\newblock It's not the algorithm, it's the data.
\newblock Communications of the ACM \textbf{60}(2)  21--23

\bibitem{cybercourt}
Fish, T.:
\newblock {AI shock: China unveils 'cyber court' complete with AI judges and
  verdicts via chat app}.
\newblock express.co.uk (2019)

\bibitem{estonia}
Niiler, E.:
\newblock {Can AI Be a Fair Judge in Court? Estonia Thinks So}.
\newblock wired.com (2019)

\bibitem{osipov2019neural}
Osipov, V., Zhukova, N., Miloserdov, D.:
\newblock Neural network associative forecasting of demand for goods.
\newblock In: Experimental Economics and Machine Learning (EEML). (2019)

\bibitem{kadyrov2019attribution}
Kadyrov, T., Ignatov, D.I.:
\newblock Attribution of customers’ actions based on machine learning
  approach.
\newblock (2019)

\bibitem{giftexchange}
Tagiew, R., Ignatov, D.:
\newblock Gift ratios in laboratory experiments.
\newblock In: Experimental Economics and Machine Learning. Volume 1627., CEUR
  Workshop Proceedings (2016)  82--93

\bibitem{jansen2019smart}
Jansen, M., Hdhili, F., Gouiaa, R., Qasem, Z.:
\newblock Do smart contract languages need to be turing complete?
\newblock In: International Congress on Blockchain and Applications, Springer
  (2019)  19--26

\bibitem{pellfirst}
Pell, B.:
\newblock Metagame: a new challenge for games and learning.
\newblock In {van den Herik}, H., Allis, L., eds.: Heuristic programming in
  artificial intelligence 3--the third computer olympiad, Ellis-Horwood (1992)

\bibitem{tagiew2008multi}
Tagiew, R.:
\newblock Multi-agent petri-games.
\newblock In: 2008 International Conference on Computational Intelligence for
  Modelling Control \& Automation, IEEE (2008)  130--135

\bibitem{saffidine2014game}
Saffidine, A.:
\newblock The game description language is turing complete.
\newblock IEEE Transactions on Computational Intelligence and AI in Games
  \textbf{6}(4) (2014)  320--324

\bibitem{zTree}
Fischbacher, U.:
\newblock z-{T}ree: Zurich toolbox for ready-made economic experiments.
\newblock Experimental Economics \textbf{10} (2007)  171--178

\bibitem{chatterjee2019probabilistic}
Chatterjee, K., Goharshady, A.K., Pourdamghani, A.:
\newblock Probabilistic smart contracts: Secure randomness on the blockchain.
\newblock In: 2019 IEEE International Conference on Blockchain and
  Cryptocurrency (ICBC), IEEE (2019)  403--412

\bibitem{ukraine}
Vasina, O., Prentice, A., King, L.:
\newblock Ukrainian ministry carries out first blockchain transactions.
\newblock reuters.com (2017)

\bibitem{herodotus}
Cassady, R.:
\newblock Auctions and Auctioneering.
\newblock University of California Press (1967)

\bibitem{friedrich}
Michael, F., Ignatov, D.I.:
\newblock General game playing b-to-b price negotiations.
\newblock In: Proceedings of the Fifth Workshop on Experimental Economics and
  Machine Learning ({EEML 2019}) co-located with the 7th International
  Conference on Applied Research in Economics ({iCare7 2019}). Volume 2479 of
  {CEUR} Workshop Proceedings., Perm, Russia, CEUR-WS.org (2019)

\bibitem{multiattribute}
Huijun, F.:
\newblock The application of multi-objective optimization to bid decision.
\newblock Systems Engineering --- Theory \& Practice \textbf{12} (1994)

\bibitem{tagiew2017}
Tagiew, R., Ignatov, D.:
\newblock Behavior mining in h-index ranking game.
\newblock In: Experimental Economics and Machine Learning. Number 1969 in CEUR
  Workshop Proceedings (2017)  52--61

\bibitem{attenineq}
Zhu, L., Lerman, K.:
\newblock Attention inequality in social media.
\newblock (2016)

\bibitem{missingwhitewoman}
Sommers, Z.:
\newblock Missing white woman syndrome: An empirical analysis of race and
  gender disparities in online news coverage of missing persons.
\newblock Journal of Criminal Law \& Criminology \textbf{106}(2) (2016)
  275--314

\bibitem{hierarchyofdeath}
Greenslade, R.:
\newblock A hierarchy of death.
\newblock The Guardian (2007)

\bibitem{attengoog}
McCurley, K.S.:
\newblock Income inequality in the attention economy.
\newblock Google Reaserch (2008)

\bibitem{matthew}
Larivière, V., Gingras, Y.:
\newblock The impact factor's matthew effect: A natural experiment in
  bibliometrics.
\newblock Journal of the Association for Information Science and Technology
  \textbf{61}(2) (2010)  424--427

\bibitem{hankey2018data}
Hankey, S., Marrison, J., Naik, R.:
\newblock Data and democracy in the digital age (2018)

\end{thebibliography}
\end{document}